\documentstyle[12pt]{article} 
\begin{document}
\vspace{1cm}
\title{\Large A Start-Timing Detector for the Collider Experiment PHENIX 
at RHIC-BNL}
\author{\large K.~Ikematsu, Y.~Iwata, K.~Kaimi$^{\ddagger}$, M.~Kaneta, T.~Kohama, \\
N.~Maeda$^{a}$, K.~Matsukado, H.~Ohnishi, K.~Ono, A.~Sakaguchi$^{b}$, \\
T.~Sugitate\footnote{Corresponding author; sugitate@hiroh2.hepl.hiroshima-u.ac.jp}, 
Y.~Sumi, Y.~Takata, M.~Tanabe$^{c}$ and A.~Yokoro}
\date{}

\maketitle
\begin{center}
\noindent
{\large\it Department of Physics, Hiroshima University, \\
Kagamiyama, Higashi-Hiroshima 739-8526, Japan}

\vspace{1cm}
\begin{abstract}
\noindent
We describe a start-timing detector for the PHENIX experiment at the 
relativistic heavy-ion collider RHIC. The role of the detector is to 
detect a nuclear collision, provide precise time information with an 
accuracy of 50ps, and determine the collision point along the beam 
direction with a resolution of a few cm. Technical challenges are that 
the detector must be operational in a wide particle-multiplicity range 
in a high radiation environment and a strong magnetic field. We present 
the performance of the prototype and discuss the final design of the 
detector.

\end{abstract}
\end{center}

\newpage
\section{Introduction}
The PHENIX (Pioneering High Energy Nuclear and Ion eXperiment) spectrometer 
\cite{1} of the Relativistic Heavy Ion Collider (hereafter abbreviated 
as RHIC) at the Brookhaven National Laboratory is now under construction. 
Phenix is composed of various sub-detectors for hadron-, lepton- and 
photon-measurements in ultra-relativistic heavy nucleus-nucleus collisions 
at $\sqrt{s}$=200 GeV/{\it c} per nucleon, as well as in proton on proton 
reactions at $\sqrt{s}$=500 GeV/{\it c}. For particle identification of 
interesting species and background rejection, time-of-flight (hereafter 
abbreviated as TOF) measurements are the crucial function. The RHIC beam 
bunches have a large longitudinal dimension, which is expected to be about 
25cm rms in the present accelerator design. It turns out that a nuclear 
collision could happen at any place in the colliding region and at any 
moment during the beam bunch crossing, which continues for about 2ns. 
For precise TOF measurements with detectors providing stop timing signals 
with 100ps resolution, we have to determine the start time very accurately. 
A detector sub-system to measure the start timing is named {\it Beam/Beam 
Counter} (abbreviated as BBC hereafter) in PHENIX.

A pair of BBC will detect secondary particles from every collision at very 
forward angles between $2.4^{\circ}$ and $5.7^{\circ}$ in the north- and 
south-ends of the central spectrometer along the beam axis. Each measures 
the arrival time of a leading charged particle, which travels at nearly 
the speed of light. An average and a difference between time measured at 
each end provide the start time measured from a common clock and the 
longitudinal vertex position along the beam axis, respectively. The vertex 
position will be employed in the first level trigger to clean up events 
and as a starting point for further online tracking analysis. 

We have completed design studies of the BBC consisting of an array of 64 
identical detector elements at each end. We have tested prototype detector 
elements with pion beams, studied performance of a small array of the 
prototypes in heavy-ion collisions at 15 GeV/{\it c} per nucleon, and have 
recently carried out quality assurance tests with the detector elements 
of the production phase. 

In section 2, we describe requirements, conceptual design and the 
technological choice for the detector. Results from a series of beam tests 
with the prototypes are given in section 3. In section 4, we discuss on our 
technological choice and optimization. We also demonstrate the start time 
determination with multiple measurements of reaction products from nuclear 
collisions and show the results from the QA tests. A summary is given in 
section 5. 

\section{Design of Detector Element and Array}
Fig. 1 shows a cut-away view of the PHENIX spectrometer. The BBC is 
positioned between the central spectrometer magnet and the muon sub-detector, 
and covers a pseudo-rapidity region from 3.0 to 3.9 over full azimuth. 
Another identical set is installed in the opposite side. Secondary 
particles emitted in this kinematical region pass through an axial hole 
of the main spectrometer magnet and go into the BBC. All the strongly 
interacting particles will be finally absorbed in the piston magnet 
placed immediately behind the BBC. 

The BBC must satisfy the following requirements: It has to primarily 
provide a precise time information of each collision in various environments 
of particle multiplicity. In case of p+p reactions, one would expect only 
a few charged particles in this acceptance. For a nuclear collision such 
as a central Au+Au collision, however, one would expect more than 1,000 
charged particles hitting the BBC. The detector has to be segmented to 
possess capability to handle the wide dynamic range of particle multiplicity 
as the entire system, and each detector element has to be well designed 
for multi-particle hits. 

Since the BBC will be placed at a very forward angle, it will be irradiated 
with an enormous amount of charged and neutral particles from beam-beam 
collisions, albedo particles from materials surrounding the detector, 
and more seriously with beam associated backgrounds. A simulation estimates 
that the gamma- and neutron-flux at the position might be about $10^{10}$ 
and $10^{11}$/cm$^{2}$/RHIC-year, respectively. The detector must be 
resistant against such a large radiation dose. In addition to this, due to 
geometrical restrictions of the PHENIX configuration, access to the BBC 
after the rolling in is very limited. The detector has accordingly to 
work maintenance free for a long period. 

There is a magnetic field of about 0.3T mostly parallel to the beam axis 
under the normal magnet operation. The detector must operate under such 
a high magnetic field, since the field is too strong to shield with any 
magnetic materials down to an operational level of standard photomultiplier 
tubes. 

Our technological choice is to build the BBC as an array detector composed 
by 64 identical Cherenkov counters. Figs 2a and 2b depict the individual 
detector element and the array mounted on a mechanical frame. The detector 
element consists of four components; a Cherenkov radiation material, 
a photomultiplier tube, a high-voltage divider module and mechanical 
accessories to pack and mount them on the frame. The Cherenkov radiator 
and the photomultiplier tube are made to be in one body, in order to avoid 
using any glue or optical grease to join the two parts together. 
The Cherenkov radiator component, which is in fact a very thick window of 
the photomultiplier tube, is made of fused Quartz and has a hexagonal shape 
inscribing 1" diameter circle and 30mm long. The photomultiplier tube was 
designed on the basis of an existing tube of 1" in diameter, Hamamatsu 
R3432, which has 15 layers of fine mesh dynode and is supposed to be 
operational in a magnetic field. Several prototypes of photomultiplier 
elements were produced in Hamamatsu Photonics and tested with pion beams 
at KEK and with heavy ion beams at AGS-BNL. 

\section{Results from Prototype Tests}
\subsection{Prototype tests with pion beams}
Three rounds (T285, T298 and T322) of beam tests were carried out at the 
12GeV Proton Synchrotron facility of KEK. A typical setup at the beam 
line T2 is illustrated in Fig. 3. We used a negative pion beam with its 
momentum of 1.6 GeV/{\it c}. A typical beam rate was around 1,000/cm$^{2}$ 
in a 1.6s beam spill with a beam definition of;
\[ DEF1 \otimes 
DEF2 \otimes 
DEF3 \otimes 
DEF4 \otimes 
REF1 \otimes 
REF2, \]
which was used as a trigger for data taking. A 35cm-long solenoid magnet 
with an inner bore radius of 10cm was employed to generate a magnetic field. 
The prototype detector element was placed at the center, where the magnet 
produces almost a flat field within 0.2\%/cm along the axis. 

ADC and TDC data from the prototype detector labeled TEST and two time 
reference counters, REF1 and REF2, were read out with standard CAMAC modules 
by a PC based DAQ system. The ADC spectrum of the prototype detector and the 
TOF spectrum between the prototype and one of reference counters are shown 
in Figs. 4a and 4b, respectively. The ADC spectrum shows a single peak, 
however, that can not be approximated by either a Poisson or a Gaussian 
distribution. 

The intrinsic timing resolution of 100ps is achieved under the magnetic 
field of 0.3T by using a leading edge discriminator without any software 
cuts or corrections. It is improved up to 50ps with the standard slewing 
correction \cite{2} using the ADC information without any cuts. The typical 
TOF spectrum after the correction is seen in Fig. 4c. Dependence of the 
current gain and the intrinsic timing resolution were studied as a function 
of strength of applied magnetic field in Fig. 5. There is a clear drop of 
ADC gain by 30\% immediately after the magnetic field was switched on, 
but shows a plateau until 0.3T and again decreases gradually beyond this 
field strength. The timing resolution after corrections, however, does not 
indicate any degradation and keeps the performance at the same level. 

\subsection{Prototype test with heavy-ion beam}
The performance of the prototype detector element for single particles 
was studied using pion beams, but in reality we also need to examine 
characteristics of the performances for a multi-particle shot in a detector 
array. For this purpose, another test was carried out using heavy-ion beams 
from the AGS at BNL. The test setup in Fig. 6 was placed downstream of the 
MPS experiment on the A1 line, and measurements were performed with 
a gold beam at 14.5 GeV/{\it c} per nucleon impinging on a 5 mm thick lead 
target. The beam envelope was defined by a beam defining counter BDF. 
Two time reference counters, REF1 and REF2 were placed in front of the 
target. They provide the start time signals for TOF measurement and measure 
the charge of the incident heavy ion. A nuclear reaction was detected with 
a small pulse height in a beam-veto counter VETO and with large pulse 
heights in tagging counters, TAG1 and TAG2, which were placed in front of 
the test detectors. Seven prototype detector elements were assembled to 
form an array with the realistic gap spacing. ADC and TDC data from all 
the counters were read out with standard CAMAC modules by a PC based DAQ 
system. Data were accumulated with a trigger of;
\[ BDF \otimes 
REF1 \otimes 
REF2 \otimes 
TAG1 \otimes 
TAG2 \otimes 
\overline{VETO} \otimes 
[\sum_{i=1}^{7}TEST_{i}]. \]
The array was set at 15$^{\circ}$ with respect to the beam line, and the 
distance from the target to the detectors was typically 35cm. We could expect 
around 10 charged particles entering the array from a central collision. 

Fig. 7a shows an ADC distribution of one of the prototype detector positioned 
at the center of the array, and Fig. 7b shows the TOF spectrum between the 
detector and one of the time reference counters. The ADC spectrum is quite 
different from those of the pion beam tests in Fig. 4a. It implies a large 
amount of background particles hitting the detector with small pulse heights 
and they might hide the ADC peak for a single particle hit. It was found 
that the majority of background particles came from adjacent detectors. 
The same slewing correction process applied to the pion beam data, was 
employed to this data. The timing resolution of 54ps was achieved in 
Fig. 7c, but the resolution before the correction is worse than that of 
pion beam tests, because of the much broader ADC spectrum.
 
\section{Discussion}
\subsection{Detector concept}
In order to accept leading particles traveling at the speed of light, 
the BBC must be set in a small space at the very forward angle. Here 
the particle multiplicity per unit area could be extremely high for 
central nuclear collisions. The multiplicity, however, varies greatly 
depending on the impact parameter of the collision in question. 
Consequently, the number of particles hitting individual detector elements 
varies, and the detector elements must accommodate multiple particles 
entering anywhere on the face of the detector. 

Once we consider a possibility of detector coupled with a photon counter, 
it is a natural solution to employ Cherenkov radiation light instead of 
scintillation light. A fast charged particle passing through a transparent 
material will radiate Cherenkov light. A photon counter optically coupled 
to this material will measures the light. Such Cherenkov light is always 
produced, but is usually negligibly small compared to the light signal 
from a scintillating material. However, if multiple particles enter a single 
detector, the Cherenkov light becomes large enough to fire a discriminator 
and trigger a TDC a few hundred ps prior to the scintillation light, 
which comes later due to the larger time constant of the scintillating 
process. This timing gap prevents precise timing measurements.

Another difficulty of precise time measurements in high multiplicity 
environments is that the photon counter should be placed at a position 
where the photon counter can view the entire detection volume at a constant 
distance. Otherwise the photon counter measures the earliest time from light 
generated in the nearest hit spot. We would then see considerable time spread 
originating from variation of the optical path length inside the detector. 
In other words, a detector designed like a paddle counter even viewed with 
double photon counters at the both ends would not work well in this case. 

The solution that we chose is a compact Cherenkov detector with a high-speed 
photomultiplier tube immediately behind the radiator material along the 
particle direction, and multiple detector elements form an array surrounding 
a beam pipe at very forward angles. 

\subsection{Radiator shape optimization}
The Cherenkov radiator shape must be optimized to collect the maximum number 
of photons. We have two parameters to be determined; the radiator length 
determining the number of photons produced, and the outer shape determining 
the photon collection by internal reflection onto the cathode plane. 
The speed of light in a Quartz radiator is slower than the speed of light 
in vacuum by a factor of the index of refraction. The path length of 
photons traversing the radiator is much longer than that of the particle, 
since the photons are radiated at the Cherenkov angle of about $45^{\circ}$ 
in Quartz with respect to the direction of the particle. As the result, 
Cherenkov photons produced near the front of radiator arrive at the 
photo-cathode later than the particle, whereas the arrival time of photons 
produced at the back of radiator is almost the same as that of the particle. 
This fact introduces a spread ($\sigma_{1}$) of photon arrival time and 
the spread results in deterioration of timing resolution of the detector. 
To minimize the spread, the length of the radiator must be as short as 
possible. On the other hand, a reasonable number of photons in a 
photomultiplier tube is necessary for a good signal ratio to noise and for 
reduction of another time spread ($\sigma_{2}$), which is originated from 
fluctuation of traveling time by photo-electrons from the 
cathode to the anodes in a photomultiplier tube. A larger number of 
photo-electrons reduces the fluctuation smaller. This temporal fluctuation 
is characterized by the transit time spread for a single photon detection, 
which is measured to be 0.33ns for this particular photomultiplier tube. 

In Fig. 8, we show the results of a simple model calculation with an 
assumption of the photo-cathode sensitivity $N_{0}$ of 100, for several 
values of the mixing parameter {\it c}, 
where $\sigma=\sqrt{c \times \sigma_{1}^{2} + \sigma_{2}^{2}}$. 
If we take $c=1$, the optimized radiator length is around 15mm. In order to 
verify the calculation and to study the effects of outer shape of radiator, 
we have tested several prototype counters with hexagonal and circular 
shaped radiators of different lengths. In Fig. 9, we plotted the test 
results of the timing resolution as a function of the radiator length. The 
model calculation with the parameter $c \approx 0.25$ seems to explain 
the global tendency rather well. Though the model is too simple to explain 
everything, we could understand the trend and choose the length to be 3cm. 
The shortest possible radiator is used, because a shorter radiator prevents 
cross talk between adjacent detector elements when a particle enters with 
a shallow angle. 

The ADC spectrum taken with a circular shaped radiator is shown in 
Fig. 10. Comparing to Fig. 4a, which was taken with a hexagonal radiator, 
there is a clear difference in the distribution. A Monte Carlo study 
simulating internal reflection of photons inside the Quartz radiator 
block explains the dependence of the spectrum shape on radiator shape 
and length very well. It is because the cathode sensitive area is smaller 
than the radiator cross section, and some of the photons can not reach 
to the sensitive area by internal reflection. The probability of this 
loss is a function of the radial position of the incident particle, 
and the function strongly depends on the radiator shape. We therefore 
see pulse height dependence on radial hit position in the ADC spectrum. 
There is no noticeable difference on the timing resolution between the 
circular and hexagonal radiators. We have chosen the hexagonal shape, 
because the higher hermeticity is important for proton on proton collisions 
and because of the preferable single peak feature in the pulse height 
distribution. 

\subsection{Radiation hardness}
Radiation damage in a photomultiplier tube usually results in a gradual 
lowering of transparency of the window material due to color center 
creation. Negligibly small effects on the photo-cathode or dynode 
materials are expected because of their quite low thickness \cite{3}. 
In this design, the photomultiplier tube has a very thick window compared 
to a standard tube and high transparencies at short wavelengths are 
essential to collect the Cherenkov light. The optical degradation of 
the window material from radiation damage is seriously examined. 
As described in the previous section, since this thick window serves 
as a Cherenkov radiator, there is no glue or grease used. This ensures 
no loss of photons at any joint gap and highly improves the reliability 
of the detector element both mechanically and in radiation hardness, 
since glues and grease lose transparency at short wavelength after 
radiation exposure. We chose fused Quartz of SPRASIL grade as the window 
material, since it is one of the most radiation hard materials and 
provides excellent transparency at short wavelength. It has been 
studied \cite{3} that, for wavelengths down to around 220nm, the Quartz 
does not indicate any serious degradation of transparency for thermal 
neutron radiation up to $10^{14}$/cm$^{2}$, which corresponds to the 
total amount of neutron radiation during 1000-year operation at RHIC. 
We tested radiation hardness of the prototype detector against 
$\gamma$-rays. The detector was irradiated with 1MR of $\gamma$-rays 
from $^{60}Co$ at the Radioisotope Center of Hiroshima University. 
The total amount of $\gamma$-ray irradiation is approximately 
$10^{5}$ times larger than that expected in 1-year RHIC operation, 
and no deterioration of the detector was observed. 

\subsection{Collision time determination}
The purpose of the BBC is to determine a unique value of the collision 
time from multiple measurements of secondary particles. This function 
is demonstrated with a small array of prototype detectors in the 
heavy-ion beam test. There are two simple ideas to derive the collision 
time. One is to adopt the smallest TDC value in each event and neglect 
any other signals. It corresponds to looking at the fastest particle 
in the collision of interest. Another is to take an average over all 
the time information in each event. This may include undesirable 
particles at the velocities slower than the speed of light. These two 
simple methods are demonstrated in Fig. 11a and 11b, where TOF spectra 
between the BBC and a time reference counter REF1 are plotted. 
For comparison, Fig. 11c shows a TOF spectrum of all the time information 
available over all events in question. Both of the methods provide the 
time resolution of around 40ps, which is slightly better than the 
timing resolution of each detector element. Although the number of 
sample measurements is only seven, it is because of multiple measurements 
of the single value in each event. In the real experiment, a larger 
number of independent measurements are expected and a better timing 
resolution can be obtained accordingly.  

\subsection{Quality assurance test}
The quality assurance test (T395) using the real detector elements built 
in the first-production-year was carried out at the 12 GeV Proton Synchrotron 
facility of KEK. The setup at the beam line T2 is almost the same as that in 
fig. 3. Fig. 12 shows results of photomultiplier current gain and timing 
resolution of all the 37 detector elements arranged in order of the gain. 
The data were taken by applying a constant high-voltage value of 2.3kV to 
each photomultiplier tube under a magnetic field of 0.3T. Although the spread 
of current gain between the lowest- and highest-gain tubes at this voltage 
extends over a factor of 4, very stable operation of all detector elements 
was observed. The timing performance is better than with the prototype 
detectors, demonstrating the uniformity of the detector production process. 

\section{Conclusion}
We have designed a start-timing detector for the PHENIX experiment at 
RHIC-BNL. The detector is composed of 64 identical detector elements, 
placed at very forward angles in the opposite ends of the central 
spectrometer. The detector element is a Cherenkov counter consisting 
of a 3cm long hexagonal shaped Quartz radiator and a photomultiplier 
tube of 1" in diameter with 12 stage mesh dynodes. The radiator and 
the tube are united into one body. 

The performance of prototype detectors was studied with pion beams 
under magnetic fields and with heavy-ion beams. The intrinsic timing 
resolution of the prototypes was about 100ps without any corrections, 
and was about 50ps with standard slewing corrections. The test experiment 
using heavy-ion beams demonstrates good performance in the collision 
time determination by an array of seven prototype detectors. The collision 
time resolution is about 40ps with this small detector array, so a better 
timing resolution is expected for events with larger particle multiplicity 
in the entire BBC configuration at the PHENIX experiment. 

The quality assurance test was performed with 37 real detector elements 
built in the first-production-year. The results of timing resolution are 
very stable, and all of the elements attain a timing resolution of 
80 - 90ps without any corrections. This 10ps level fluctuation is due 
to the difference in current gain of the photomultiplier tubes. After 
slewing corrections, a resolution as good as 30ps is obtained, 
independent of the current gain. 

It is worth noting that, though the detector was developed for a particular 
experiment, the detector element, which is now commercially available 
as R6178 from Hamamatsu Photonics, can be employed in any experimental 
field requiring a fine granularity of 1" in space and an excellent timing 
resolution better than 50ps under a magnetic field. 

\section*{Acknowledgements}
This study has been motivated for the PHENIX experiment at BNL-RHIC, and the 
authors are most grateful to the entire PHENIX collaboration for invaluable 
discussions during the studies and in particular for the permission to use 
the picture of Fig.1. The authors also express their sincere thanks to Prof. 
J.~Chiba and the staff of the KEK 12GeV PS facility for their excellent work 
and support. They are also indebted to Dr. K.~Foley and Dr. E.~Asher for their 
kind offer sharing the heavy-ion beam at MPS-AGS. Their gratitude is also 
extended to the MPS technical staffs for their invaluable support. Data 
analysis was primarily carried out at the Data-Analysis Laboratory for 
High-Energy Physics, Hiroshima University. This work is supported by a 
Grant-in-Aid for Scientific Research (05452029) by the Ministry of 
Education, Science, Sports and Culture, Japan and in part by the 
U.S.-Japan High Energy Physics Collaboration Treaty. \\

\vspace {5cm}
\noindent
$^{\ddagger}$ {deceased}\\
$^{a}$ {present address: Physics Department, Florida State University, 
                         Tallahassee, FL, USA}\\
$^{b}$ {present address: Physics Department, Osaka University, Osaka, Japan}\\
$^{c}$ {present address: Nippon Mikuniya Co., Tokyo, Japan}\\

\newpage

\begin{center}
    {\Large \bf Figure Captions}
\end{center}
\begin{description}
\item[Fig.1]  A cut-away view of the PHENIX spectrometer at RHIC-BNL. The BBC 
is depicted surrounding the beam pipe between the central magnet and the muon 
magnet. There is no access way to the BBC after rolling-in of the spectrometer.

\item[Fig.2]  (a) Components of the detector element; a photomultiplier tube 
of 1" in diameter with a 30mm thick Quartz window, a breeder circuit, an 
aluminum attachment glued on the top of window, a sleeve of 0.2mm-thick 
stainless steel and a plastic sleeve insert. (b) The detector elements mounted 
on the frame of 30cm in diameter and 25cm long. The frame can be split into 
two parts at the middle plane in order to put a 3" beam pipe in the 10cm 
diameter hole at the center. 

\item[Fig.3]  A typical setup view of the KEK tests with pion beams. Not 
in scale. 

\item[Fig.4]  (a) ADC distribution of a prototype detector with pion beams. 
TOF spectra between the prototype detector and REF1 (b) before and (c) after 
the slewing corrections.

\item[Fig.5]  The ADC peak channel and the intrinsic timing resolution of a 
prototype detector with pion beams, as a function of the strength of applied 
magnetic field.

\item[Fig.6]  A setup view of the prototype array tests with heavy-ion beams 
at AGS. Not in scale. 

\item[Fig.7]  (a) ADC distribution of a prototype detector positioned at the 
center of array. TOF spectra between the detector and REF1 (b) before and 
(c) after slewing corrections. 

\item[Fig.8]  Timing resolution expected in a model calculation as a function 
of radiator length. 

\item[Fig.9]  The test results of intrinsic timing resolution and ADC peak 
channel as a function of radiator length. Measured with pion beams. 

\item[Fig.10]  ADC distribution measured by a prototype detector with a 
circular shape radiator. The distribution is quite different from that of 
a hexagonal shaped radiator in fig. 4a. 

\item[Fig.11]  Demonstration of the collision time determination in heavy-ion 
collisions with an array of seven prototype detectors. (a) TOF spectrum 
between REF1 and a detector measured the smallest TDC value in each event. 
(b) TOF spectrum between REF1 and an average of all the valid TDC values 
measured in each event. (C) TOF spectrum between REF1 and all the 
detectors measured valid TDC values over the events. 
The top two spectra contain 13k events and the bottom one contains 
58k hits. The average hit-multiplicity is 4.6. They show the timing 
resolution of 43ps, 37ps and 48ps, respectively. 

\item[Fig.12]  The results of quality assurance test performed with pion beams 
at KEK. 37 detector elements constructed in the first-year-production-phase 
are tested for the timing resolution and the current gain. 

\end{description}
\end{document}